\documentclass[conference]{IEEEtran}
\IEEEoverridecommandlockouts

\usepackage{textcomp}
\usepackage{xcolor}

\usepackage{amsmath,amssymb,amsfonts}
\usepackage{algorithm}
\usepackage{algorithmic}
\usepackage{graphicx}
\graphicspath{ {Figures/} }
\usepackage{stfloats}
\usepackage{multicol}
\usepackage{multirow}
\usepackage{tablefootnote}
\usepackage[flushleft]{threeparttable}
\usepackage{color, colortbl}
\usepackage{lipsum}

\usepackage{hyperref}
\usepackage{balance}
\usepackage{cite}

\newcommand{\Fp}{\mathbb{F}_p}

\newcommand{\Fpa}{\mathbb{F}_{p^{2}}}

\newcommand{\Fpd}{\mathbb{F}_{p^{12}}}
\newcommand{\Fpk}{\mathbb{F}_{p^{k}}}

\newcommand{\Zq}{\mathbb{Z}_q}
\newcommand{\Zqn}{\mathbb{Z}_q^n}

\newcommand{\GLn}{\mathbb{GL}_n}
\newcommand{\Ga}{\mathbb{G}_1}
\newcommand{\Gb}{\mathbb{G}_2}
\newcommand{\Gt}{\mathbb{G}_{T}}

\definecolor{Gray}{gray}{0.9}
\definecolor{Highlight}{rgb}{0.99, 0.99, 0.9}

\def\BibTeX{{\rm B\kern-.05em{\sc i\kern-.025em b}\kern-.08em
    T\kern-.1667em\lower.7ex\hbox{E}\kern-.125emX}}
\begin{document}

\title{
Privacy-Preserving Edge Computing from Pairing-Based Inner Product Functional Encryption
\thanks{\textcopyright $\,$ 2023 IEEE. Personal use of this material is permitted. Permission from IEEE must be obtained for all other uses, in any current or future media, including reprinting/republishing this material for advertising or promotional purposes, creating new collective works, for resale or redistribution to servers or lists, or reuse of any copyrighted component of this work in other works.}
\thanks{A revised version of this paper was published in the proceedings of the 2023 IEEE Global Communications Conference (GLOBECOM) - DOI: \href{https://dx.doi.org/10.1109/GLOBECOM54140.2023.10436785}{10.1109/GLOBECOM54140.2023.10436785}}
}

\author{
\IEEEauthorblockN{Utsav Banerjee}
\IEEEauthorblockA{Electronic Systems Engineering \\ Indian Institute of Science, Bengaluru, India}
}

\maketitle

\begin{abstract}
Pairing-based inner product functional encryption provides an efficient theoretical construction for privacy-preserving edge computing secured by widely deployed elliptic curve cryptography.
In this work, an efficient software implementation framework for pairing-based function-hiding inner product encryption (FHIPE) is presented using the recently proposed and widely adopted BLS12-381 pairing-friendly elliptic curve.
Algorithmic optimizations provide $\approx 2.6 \times$ and $\approx 3.4 \times$ speedup in FHIPE encryption and decryption respectively, and extensive performance analysis is presented using a Raspberry Pi 4B edge device.
The proposed optimizations enable this implementation framework to achieve performance and ciphertext size comparable to previous work despite being implemented on an edge device with a slower processor and supporting a curve at much higher security level with a larger prime field.
Practical privacy-preserving edge computing applications such as encrypted biomedical sensor data classification and secure wireless fingerprint-based indoor localization are also demonstrated using the proposed implementation framework.
\end{abstract}


\section{Introduction}
\label{sec:intro}

Rapid growth in cloud computing technology has led to the rise of ``computation as a service'', where big data computing tasks are outsourced to the cloud. Although the cloud infrastructure with its powerful servers enables a wide variety of computationally complex tasks, there are also serious privacy concerns regarding the data being processed. Such concerns have motivated the emerging research area of ``privacy-preserving computation'', also known as ``computation on encrypted data'', which includes a plethora of cryptographic tools \cite{shan_securecompute_2018, menezes_handbook_2018} such as functional encryption, homomorphic encryption, multi-party computation and secret sharing, which allow secure outsourced computation without revealing the data being computed upon.

Functional encryption allows the computation of a function of the encrypted data during decryption without revealing the original data \cite{boneh_fe_2011}. One of the most well-known practical constructions of such functional encryption allows computing inner products of private vectors, also known as ``inner product encryption'' \cite{kim_ipe_2018}.
Such inner product functional encryption schemes are based on bilinear map operations over pairing-friendly elliptic curves \cite{nadia_pairing_2017}.
Wide deployment of elliptic curve cryptography \cite{hankerson_ecc_2006} in today's Internet-connected systems, including Internet of Things (IoT) devices \cite{scott_ecciot_2020}, makes pairing-based inner product functional encryption naturally suitable for many privacy-preserving computing applications.
While pairing-based inner product encryption has been explored for cloud applications and implemented on server-scale or desktop-scale high-performance processors in previous literature \cite{kim_ipe_2018}, their optimized implementation for privacy-preserving edge computation is yet to be explored.

In this work, an efficient software implementation framework for pairing-based inner product functional encryption based on the recently proposed and widely adopted BLS12-381 pairing-friendly elliptic curve \cite{ietf_pairingcurves_2020} is presented.
Various algorithmic optimizations are proposed for both encryption and decryption. The proposed optimizations are validated by detailed performance analysis using measurement results obtained from a Raspberry Pi 4B edge device \cite{raspberry_pi_4b}.
Due to the algorithmic optimizations and efficient software implementation, the encryption and decryption execution times and ciphertext sizes obtained using the proposed framework are comparable to previous work \cite{kim_ipe_2018} despite being implemented on an edge device with a slower processor and supporting a curve at much higher security level with larger prime field.
Privacy-preserving edge computing applications such as biomedical sensor data classification and wireless fingerprint-based indoor localization are also demonstrated along with practical performance metrics obtained from this implementation framework.

\section{Background}
\label{sec:background}

\subsection{Elliptic Curves and Pairings}

Elliptic curves are widely used for public key cryptography, e.g., key exchange, digital signatures and authentication protocols \cite{hankerson_ecc_2006}. Certain special elliptic curves allow the computation of bilinear maps to finite fields, also known as pairings \cite{nadia_pairing_2017}, to enable novel security applications such as signature aggregation, multi-party key agreement, inner product encryption, identity-based encryption and attribute-based encryption.
Let $E : y^2 = x^3 + ax + b$ be a pairing-friendly elliptic curve defined over prime field $\Fp$. Let $\Ga$ be a cyclic subgroup of $E(\Fp)$ of order $q$. Let $\Gb$ be a cyclic subgroup of $E(\Fpk)$ of order $q$, where the embedding degree $k$ is the smallest integer such that $q \, | \, (p^k - 1)$. Let $\Gt$ be a $q$-order subgroup of the multiplicative group $\Fpk^{*}$. Then, a pairing computation is defined by the map $e : \Ga \times \Gb \rightarrow \Gt$ which satisfies bilinearity property: $e(aP, bQ) = e(P, Q)^{ab}$, where $P \in \Ga$, $Q \in \Gb$, $a, b \in \Zq$ and $\Zq$ is the quotient ring of integers modulo $q$. Here, $aP$ and $bQ$ are known as elliptic curve scalar multiplications ($a, b$ are scalars and $P, Q$ are elliptic curve points).
Various pairing-friendly elliptic curves such as BN-254, BLS12-381, BN-462, etc are supported by cryptographic software libraries, e.g., MIRACL \cite{miracl_crypto}. Since recent advances in cryptanalysis have reduced the theoretical security of BN-254, the Internet Engineering Task Force (IETF) is considering BLS12-381 and BN-462 for standardization (recommended as optimistic and conservative choices respectively) at 128-bit security level \cite{ietf_pairingcurves_2020}. BN-462 is more computationally expensive compared to BLS12-381, thus motivating the wide adoption of BLS12-381 by several applications.
Table \ref{table:pairingcurvesummary} summarizes the curve equations, prime sizes and security levels of these curves. All three curves have embedding degree $k=12$.

\begin{table}[!t]
\renewcommand{\arraystretch}{1.25}
\caption{Parameters of Commonly Used Pairing-Friendly Curves}
\label{table:pairingcurvesummary}
\centering
\begin{tabular}{|l|c|c|c|c|}
\hline
\rowcolor{Gray}
\textbf{Curve} & \textbf{Equation} & \textbf{$\lceil \log_2p \rceil$} & \textbf{$\lceil \log_2q \rceil$} & \textbf{Security Level} \\
\hline
BN-254 & $y^2 = x^3 + 2$ & 254 & 254 & $\approx$ 100-bit \\
\hline
BLS12-381 & $y^2 = x^3 + 4$ & 381 & 255 & $\approx$ 126-bit \\
\hline
BN-462 & $y^2 = x^3 + 5$ & 462 & 462 & $\approx$ 134-bit \\
\hline
\end{tabular}
\end{table}

\subsection{Function-Hiding Inner Product Encryption (FHIPE)}

Functional encryption allows the computation of a function of the encrypted data upon decryption \cite{boneh_fe_2011}.
For inner product functional encryption, the secret key and the ciphertext are respectively associated with vectors $\boldsymbol{x} \in \Zqn$ and $\boldsymbol{y} \in \Zqn$ of length $n$ with elements in $\Zq$. Decryption result in the computation of their inner product $\langle \boldsymbol{x}, \boldsymbol{y} \rangle \in \Zq$.
The ``function-hiding inner product encryption'' (FHIPE) scheme by Kim et al. \cite{kim_ipe_2018} is one of the most efficient pairing-based constructions. It also ensures that vectors $\boldsymbol{x}$ and $\boldsymbol{y}$ remain hidden from the decryptor, thus making it well-suited for many privacy-sensitive computation applications.
The FHIPE scheme of Kim et al. \cite{kim_ipe_2018} can be described using the following four algorithms:
\begin{itemize}
\item \textsf{Setup} ($1^\lambda$, $S$) $\rightarrow$ ($pp$, $msk$) $\,$ : given security parameter $\lambda$, the setup algorithm samples a matrix $\boldsymbol{B} \in \GLn(\Zq)$ and outputs public parameters $pp = (\Ga, \Gb, \Gt, q, e, S)$ for the bilinear map and master secret key $msk = (pp, G_1, G_2, \boldsymbol{B}, \boldsymbol{B}^{*})$, where $\boldsymbol{B}^{*} = det(\boldsymbol{B}) \cdot (\boldsymbol{B}^{-1})^{T}$. Here, $n$ is a positive integer, $\GLn(\Zq)$ denotes the general linear group of $n \times n$ invertible matrices over $\Zq$ and $S$ is a subset of $\Zq$ with polynomial size $|S| = poly(\lambda)$.
\item \textsf{KeyGen} ($msk$, $\boldsymbol{x}$) $\rightarrow$ $sk_{\boldsymbol{x}}$ $\,$ : the key generation algorithm outputs secret key $sk_{\boldsymbol{x}} = (k_1, \boldsymbol{k}_2) = (\alpha \cdot det(\boldsymbol{B}) \, G_1, \, \alpha \cdot \boldsymbol{x} \cdot \boldsymbol{B} \, G_1)$ corresponding to vector $\boldsymbol{x} \in \Zqn$, where $\alpha \in \Zq$ is a uniformly random element.
\item \textsf{Encrypt} ($msk$, $\boldsymbol{y}$) $\rightarrow$ $ct_{\boldsymbol{y}}$ : the encryption algorithm generates ciphertext $ct_{\boldsymbol{y}} = (c_1, \boldsymbol{c}_2) = (\beta \, G_2, \, \beta \cdot \boldsymbol{y} \cdot \boldsymbol{B}^{*} \, G_2)$ corresponding to vector $\boldsymbol{y} \in \Zqn$, where $\beta\in \Zq$ is a uniformly random element.
\item \textsf{Decrypt} ($pp$, $sk_{\boldsymbol{x}}$, $ct_{\boldsymbol{y}}$) $\rightarrow$ $z \in S \, \cup \, \{\perp\}$ : the decryption algorithm computes: \\
\hspace*{12mm} $d_1 = e(k_1, c_1) = e(G_1, G_2) ^ {\alpha \beta \cdot det(\boldsymbol{B})}$ \\
\hspace*{5mm} and, $d_2 = e(\boldsymbol{k}_2, \boldsymbol{c}_2) = e(G_1, G_2) ^ {\alpha \beta \cdot \boldsymbol{x} \cdot \boldsymbol{B} \cdot (\boldsymbol{B}^{*})^{T} \cdot \boldsymbol{y}^{T}}$ \\
Since $\boldsymbol{B} \cdot (\boldsymbol{B}^{*})^{T} = det(\boldsymbol{B}) \cdot \boldsymbol{\mathbb{I}}_{n \times n}$, where $\boldsymbol{\mathbb{I}}_{n \times n}$ is the $n \times n$ identity matrix, $d_2 = e(G_1, G_2) ^ {\alpha \beta \cdot det(\boldsymbol{B}) \cdot \langle \boldsymbol{x}, \boldsymbol{y} \rangle}$. Therefore, $d_2 = d_1 ^ {\langle \boldsymbol{x}, \boldsymbol{y} \rangle}$, and the decryptor needs to check if any $z \in S$ exists such that $d_2 = d_1^z$. The output is $z = \langle \boldsymbol{x}, \boldsymbol{y} \rangle$ if such a value can be found, otherwise the output is $\perp$.
The decryption is correct if and only if the vectors $\boldsymbol{x}, \boldsymbol{y}$ satisfy the property $\langle \boldsymbol{x}, \boldsymbol{y} \rangle \in S$.
\end{itemize}
For analyzing the computation cost of this FHIPE scheme, only \textsf{Encrypt} and \textsf{Decrypt} are considered since \textsf{Setup} and \textsf{KeyGen} need to be performed once at the beginning of application setup. Within \textsf{Encrypt} and \textsf{Decrypt}, only the operations in $\Ga, \Gb, \Gt$ are analyzed since these together account for 99\% of the computation cost \cite{banerjee_phd_2021}:
\begin{itemize}
\item \textsf{Encrypt} requires $n+1$ scalar multiplications in $\Gb$.
\item \textsf{Decrypt} requires 1 pairing for $d_1$, an $n$-fold multi-pairing (product of $n$ pairings) for $d_2$, and a bounded discrete logarithm over $\Gt$ to find $z \in S$ such that $d_2 = d_1^z$.
\end{itemize}
If the elements of $\boldsymbol{x}$ and $\boldsymbol{y}$ are bounded as $x_i \le B_x$ and $y_i \le B_y$ respectively, then $\langle \boldsymbol{x}, \boldsymbol{y} \rangle \le n B_x B_y \,\,\, \Rightarrow \,\,\, n B_x B_y < q \,\,\,\,\,\,\, \text{since} \,\,\, \langle \boldsymbol{x}, \boldsymbol{y} \rangle \in S \subset \Zq$. However, in practice, it will be required to have $n B_x B_y \ll q$ in order to efficiently compute the bounded discrete logarithm for successful decryption.
\section{FHIPE Implementation and Optimization}
\label{sec:implementation}

\begin{figure}[!t]
\centering
\includegraphics[width=3.4in]{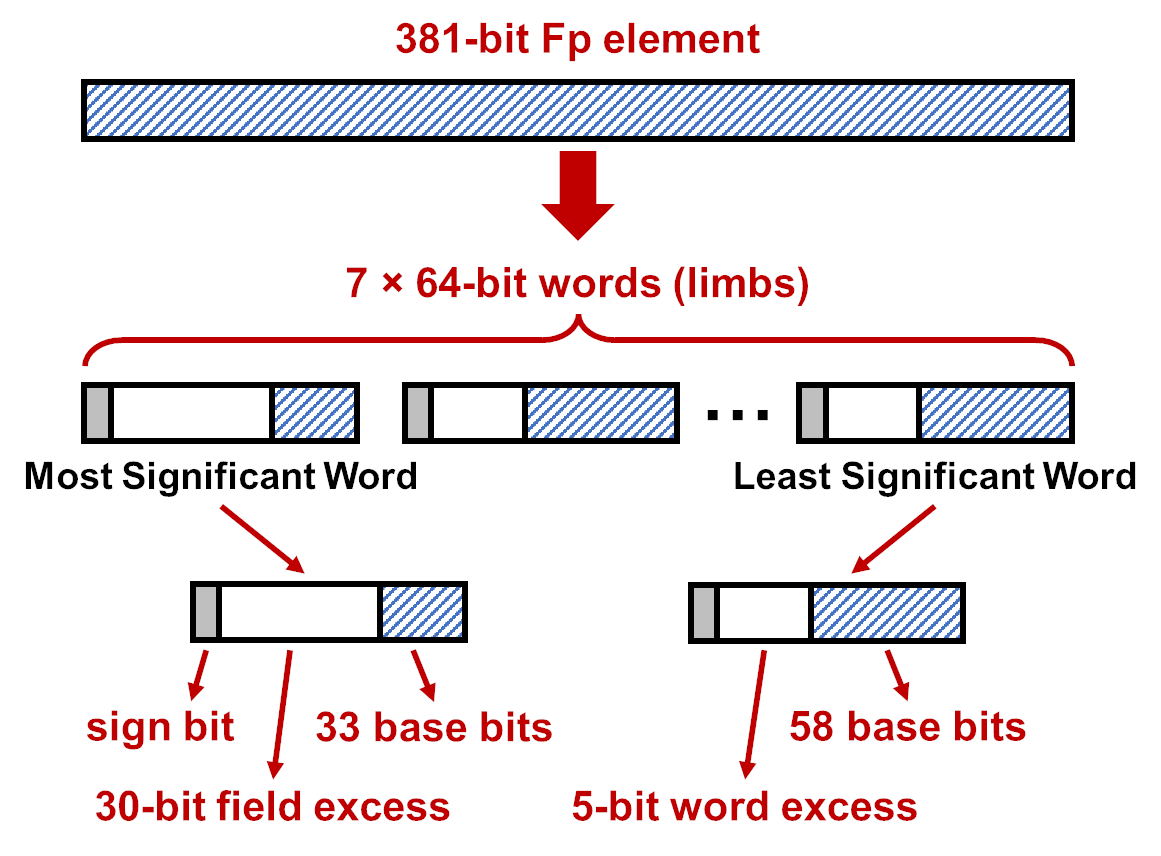}
\caption{Structure of 381-bit $\Fp$ element representation in MIRACL software.}
\label{fig:fp_structure}
\end{figure}

The FHIPE scheme is implemented in C using the elliptic curve and pairing-based cryptography library provided by MIRACL \cite{miracl_crypto}. The BLS12-381 pairing-friendly curve is used for this implementation. A Raspberry Pi 4 Model B single board computer \cite{raspberry_pi_4b} (with a 1.8~GHz quad-core ARM Cortex-A72-based Broadcom BCM2711 system-on-chip, an 8~GB LPDDR4-3200 SDRAM off-chip memory and a 128~GB microSDXC A2/V30/U3 UHS-I persistent storage) is used as the evaluation platform. In order to fully realize the potential of its 64-bit processor architecture, the 64-bit operating system version of Raspberry Pi OS (kernel version 5.15) is installed. Furthermore, the MIRACL library is configured to store the 381-bit $\Fp$ elements in the form of an array of 64-bit words (also known as limbs). As shown in Figure \ref{fig:fp_structure}, this representation contains several ``word excess'' and ``field excess'' bits to help speed up modular arithmetic computations \cite{miracl_crypto}.
The MIRACL-based software implementation is constant-time, that is, the execution time is independent of any secret inputs or parameters, and includes various state-of-the-art side-channel countermeasures.
All software code is compiled using GCC (compiler version 10.2.1 with -O3 flag) which is included with the Raspberry Pi operating system.
Next, various optimizations of the FHIPE encryption and decryption computations are discussed along with measured performance results obtained from the Raspberry Pi 4B setup.

\begin{figure}[!t]
\centering
\includegraphics[width=3.4in]{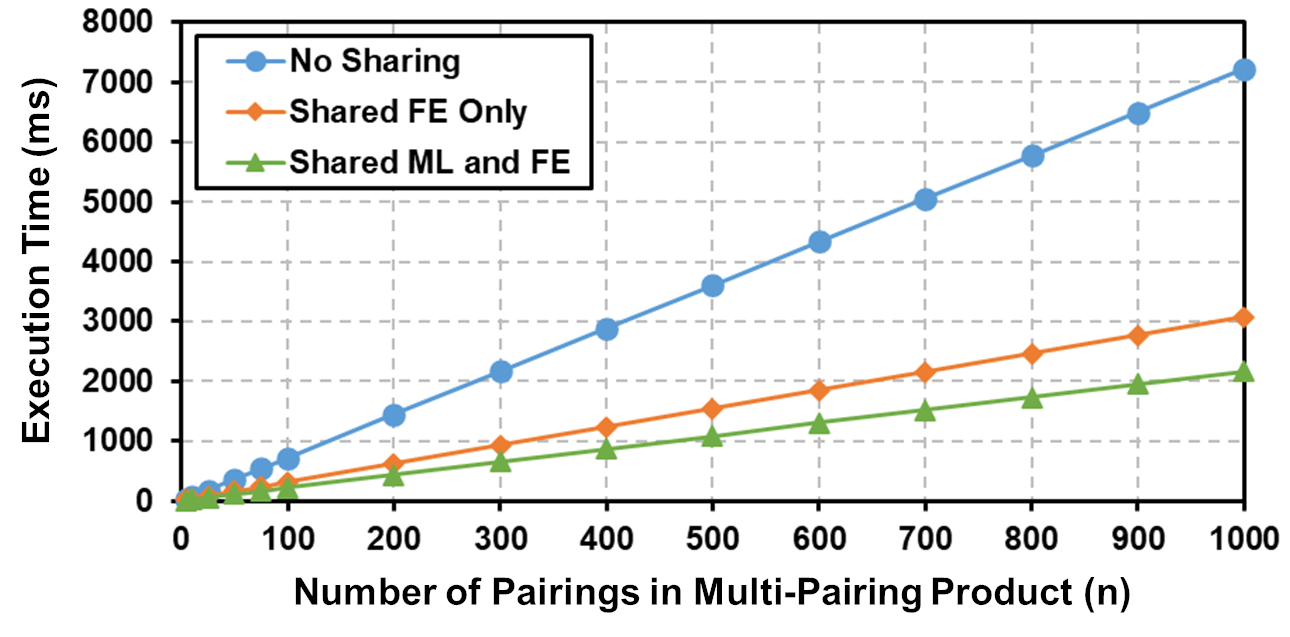}
\caption{Computation cost of $n$-fold multi-pairing with various optimizations.}
\label{fig:multi_pairing}
\end{figure}

\subsection{Scalar Multiplication in $\Gb$ for FHIPE Encryption}

Scalar multiplications in $\Gb$ account for majority of the computation cost in FHIPE \textsf{Encrypt}. As discussed earlier, for the BLS12-381 curve $E: y^2 = x^3 + 4$ with embedding degree $k = 12$, the group $\Gb$ is defined over $E(\Fpd)$. The curve also possesses an efficiently computable isomorphism $\Psi: E \rightarrow E'$ which allows mapping points from $E(\Fpd)$ to $E'(\Fpa)$, where $E': y^2 = x^3 + 4 (1 + \alpha)$ is the sextic twist curve of $E$. This helps simplify the pairing computations in $\Gb$ from $\Fpd$ to $\Fpa$ and also compresses its elements. To speed up scalar multiplications on $\Gb$, efficiently computable endomorphisms and scalar decomposition techniques are employed, as proposed by Galbraith, Lin and Scott \cite{galbraith_endomorphism_2009}, also known as the GLS method. The scalar multiple $mP$ for any point $P \in E'(\Fpa)$ and scalar $m \in \Zq$ is computed using 4-dimensional decomposition as:
\[
mP = m_0P + m_1\psi(P) + m_2\psi^2(P) + m_3\psi^3(P)
\]
Here, $\psi = \Psi \, \pi_p \, \Psi^{-1}$ (where $\pi_p$ is the $p$-power Frobenius map defined over $E(\Fpd)$) is the efficiently computable GLS endomorphism \cite{nadia_pairing_2017} such that $\psi(P) = \lambda P$, that is, it corresponds to scalar multiplication by $\lambda$. Similarly, $\psi^2(P) = \psi(\psi(P)) = \lambda^2 P$ and $\psi^3(P) = \psi(\psi(\psi(P))) = \lambda^3 P$.
Fast explicit formulas exist for computing  $\psi^i(P) = \lambda^i P$ ($i \ge 1$) and does not require explicit scalar multiplications by $\lambda^i$, thus making it very efficient.
Further, $m_0$, $m_1$, $m_2$, $m_3$ are short scalars such that $m \equiv m_0 + m_1 \lambda + m_2 \lambda^2 + m_3 \lambda_3$ (mod $q$) ensuring $mP = m_0 P + \sum_{i=1}^{3} m_i \psi^i(P)$. These scalars are computed using the method of lattice bases and optimal scalar decompositions \cite{nadia_pairing_2017} such that each $m_i$ ($i \in \{0, 1, 2, 3\}$) is much smaller than the original scalar $m$.
Then, $mP$ is computed using a multi-scalar multiplication \cite{hankerson_ecc_2006} involving bit-wise iteration over all the four small scalars $m_i$ simultaneously, as opposed to bit-wise iteration over the much larger scalar $m$. This is a memory-time trade-off and requires pre-computing and storing not only the points $\psi(P)$, $\psi^2(P)$, $\psi^3(P)$ but also a lookup table of the point sums $P + \sum_{i=1}^{3} b_i \psi^i(P)$ for all the 8 possible combinations of $b_i \in \{0, 1\}$. The MIRACL software library supports scalar re-coding and other optimization techniques \cite{bos_exponentiating_2014, faz_efficient_2015} to further speed up the 4-dimensional GLS-decomposed multi-scalar multiplication. Overall, the fast $\Gb$ scalar multiplication takes 1.62~ms on the Raspberry Pi 4B setup with 64-bit architecture described earlier. This is $\approx 2.6 \times$ faster than traditional scalar multiplication in $\Gb$ with 4-bit windows, which takes 4.16~ms on the same setup.

\subsection{Multi-Pairing in $\Ga \times \Gb \rightarrow \Gt$ for FHIPE Decryption}

Computation of multi-pairing \cite{nadia_pairing_2017}, which is essentially product of multiple pairings, is important for FHIPE \textsf{Decrypt} as it eventually leads to the inner product functionality. An $n$-fold multi-pairing has the following generalized expression:
\[
\prod_{j=1}^{n} \, e(P_j, Q_j) \, = \, e(P_1, Q_1) \times e(P_2, Q_2) \times \, \cdots \, \times e(P_n, Q_n)
\]
where $P_j \in \Ga$ and $Q_j \in \Gb$ for $j \in [1, n]$. The pairing calculation consists of two main components -- Miller Loop (ML) and Final Exponentiation (FE). It is possible to speed up multi-pairing by sharing them across all $n$ pairing instances \cite{granger_product_2006, scott_pairing_2019}. Figure \ref{fig:multi_pairing} compares the multi-pairing execution times, measured using the Raspberry Pi 4B setup described earlier, for different $n$: (1) with no shared computation, (2) sharing only FE, and (3) sharing both ML and FE. Compared to (1), up to $2.35 \times$ speedup is provided by (2), and another $\approx 30\%$ speedup by (3), that is, total $\approx 3.4 \times$ improvement.

\subsection{Bounded Discrete Logarithm in $\Gt$ for FHIPE Decryption}

In FHIPE \textsf{Decrypt}, after computing $d_1 = e(k_1, c_1)$ using pairing and $d_2 = e(\boldsymbol{k}_2, \boldsymbol{c}_2)$ using multi-pairing, another crucial component is solving the discrete logarithm $d_2 = d_1 ^ {\langle \boldsymbol{x}, \boldsymbol{y} \rangle}$ in $\Gt$ to get the final inner product $\langle \boldsymbol{x}, \boldsymbol{y} \rangle$. Since unbounded discrete logarithms over $\Zq$ are computationally intractable \cite{hankerson_ecc_2006, nadia_pairing_2017}, a bound is set in the form of $\langle \boldsymbol{x}, \boldsymbol{y} \rangle \in S \subset \Zq$ where $S = \{0, 1, \cdots, s-1\}$ so that $|S| = s$ and $nB_xB_y < s \ll q$.

\begin{figure}[!b]
\centering
\includegraphics[width=3.0in]{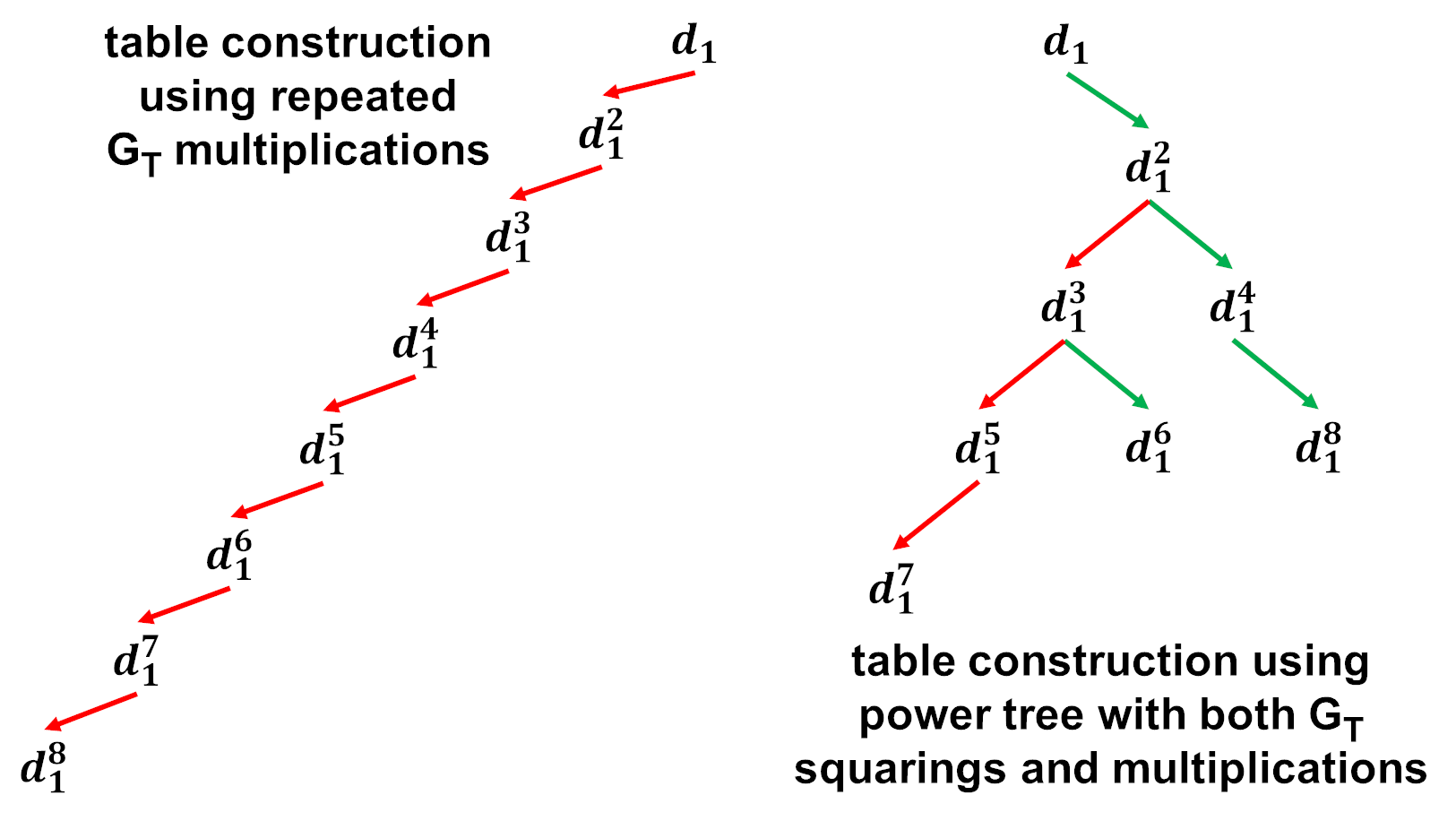}
\caption{Computation of $d_1^2, d_1^3, \cdots, d_1^8$ for $d_1 \in \Gt$ with (left) only multiplications and (right) both squarings and multiplications with power tree.}
\label{fig:power_tree}
\end{figure}

The bounded discrete logarithm is computed using the Baby-Step Giant-Step method \cite{menezes_handbook_2018} described in Algorithm \ref{algo:baby_step_giant_step}.
Here, line 2 involves the computation of a lookup table $T$ of $\Gt$ elements $d_1^j \,\,\, \forall \,\,\, j \in [0, \alpha)$, where $\alpha = \lceil \, \sqrt{s} \, \rceil$, and line 3 requires the computation of $d_1^\alpha$ followed by its inversion.
The straightforward method would involve computing $d_1^2, d_1^3, \cdots, d_1^{\alpha}$ through $\alpha - 1$ repeated field multiplications ($d_1^0$ and $d_1^1$ are trivial). However, the computation can be sped up by replacing half of the multiplications with faster Granger-Scott cyclotomic squarings \cite{nadia_pairing_2017}.
The proposed method employs Knuth's power tree \cite{knuth_comp_1997} and is illustrated in Figure \ref{fig:power_tree} for $\alpha = 8$, where red arrows denote $\Gt$ multiplications and green arrows denote faster $\Gt$ cyclotomic squarings. The power tree enables the judicious use of field squarings -- all the even powers computed with squarings and all the odd powers with multiplications of values calculated earlier in the power tree.
The number of $\Gt$ multiplications (resp. squarings) required is $\frac{\alpha}{2} - 1$ (resp. $\frac{\alpha}{2}$) and $\frac{\alpha - 1}{2}$ (resp. $\frac{\alpha - 1}{2}$) for even and odd $\alpha$ respectively.
Each $\Gt$ multiplication and cyclotomic squaring takes 20.4~$\mu$s and 10.1~$\mu$s respectively on the Raspberry Pi 4B setup described earlier. Therefore, the proposed method using power-tree is $\approx 25$\% faster than repeatedly using only multiplications.
The table lookup in line 7 is implemented as a constant-time brute-force search. While hash-table-based search is faster \cite{banerjee_phd_2021, ahsan_ants_2022}, it has prohibitively large memory overheads for large $\alpha$ and hence not implemented.
The dummy variables in line 10 are included as a modification to the original algorithm \cite{menezes_handbook_2018} to ensure that the loop is constant-time (always $\alpha$ iterations) and does not leak any side-channel information about the final result $z$.

\begin{algorithm}[!b]
\caption{Solving bounded discrete logarithm in $\Gt$ using Baby-Step Giant-Step (modified from \cite{menezes_handbook_2018})}
\label{algo:baby_step_giant_step}
\begin{algorithmic}[1]
\REQUIRE $d_1, d_2 \in \Gt$ and $S = \{0, 1, \cdots, s-1\} \subset \Zq$
\ENSURE $z \in S$ such that $d_2 = d_1^z$
\STATE $\alpha \leftarrow \lceil \, \sqrt{s} \, \rceil$
\STATE Pre-compute table $T = \{ \, (j, \, d_1^j) \, \}_{j=0}^{\alpha-1}$
\STATE $t_0 \leftarrow d_1^{-\alpha}$
\STATE $t_1 \leftarrow d_2$
\STATE $z \leftarrow \perp$
\FOR{($i = 0$; $i < \alpha$; $i = i + 1$)}
\IF{there exists $(j, \, d_1^j) \in T$ such that $t_1 = d_1^j$}
\STATE $z \leftarrow i \alpha + j$
\ELSE
\STATE $z_{dummy} \leftarrow i \alpha + j_{dummy}$
\ENDIF
\STATE $t_1 \leftarrow t_1 \cdot t_0$
\ENDFOR
\RETURN $z$
\end{algorithmic}
\end{algorithm}

\subsection{Overall Computation and Communication Costs}

\begin{figure}[!t]
\centering
\includegraphics[width=3.4in]{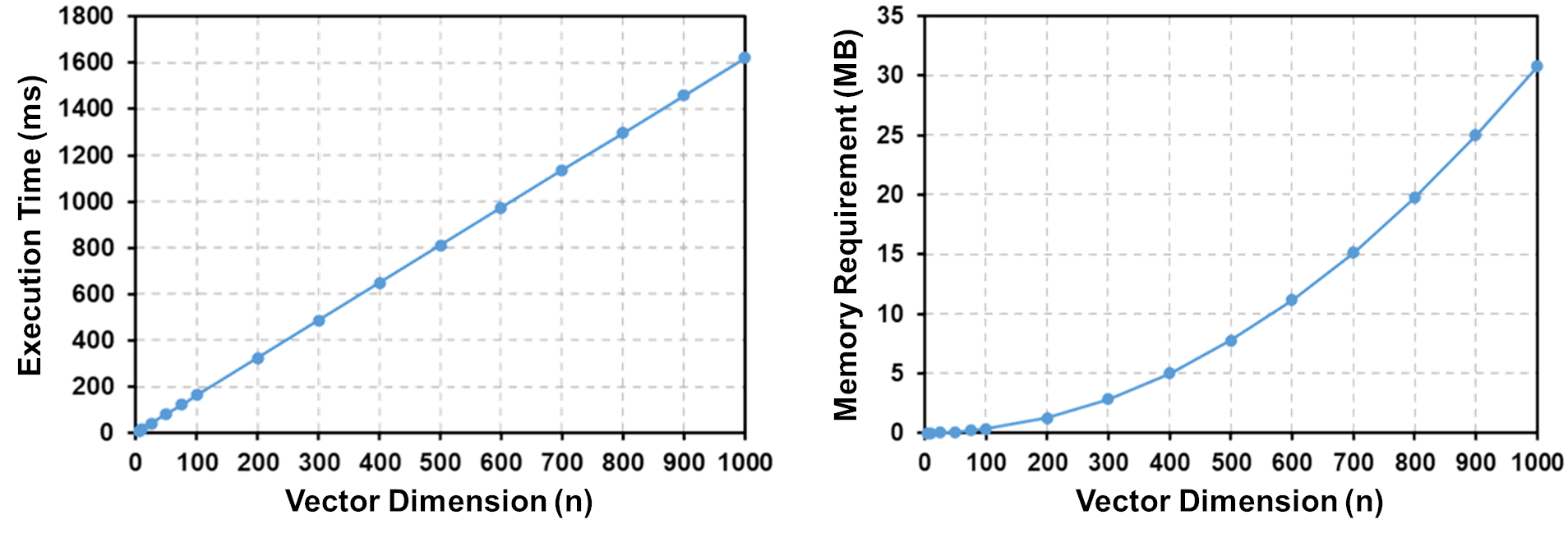}
\caption{Compute cost of FHIPE \textsf{Encrypt} for different vector dimensions ($n$): (left) execution time and (right) memory requirement.}
\label{fig:fhipe_encrypt_cost}
\end{figure}

\begin{figure}[!t]
\centering
\includegraphics[width=3.4in]{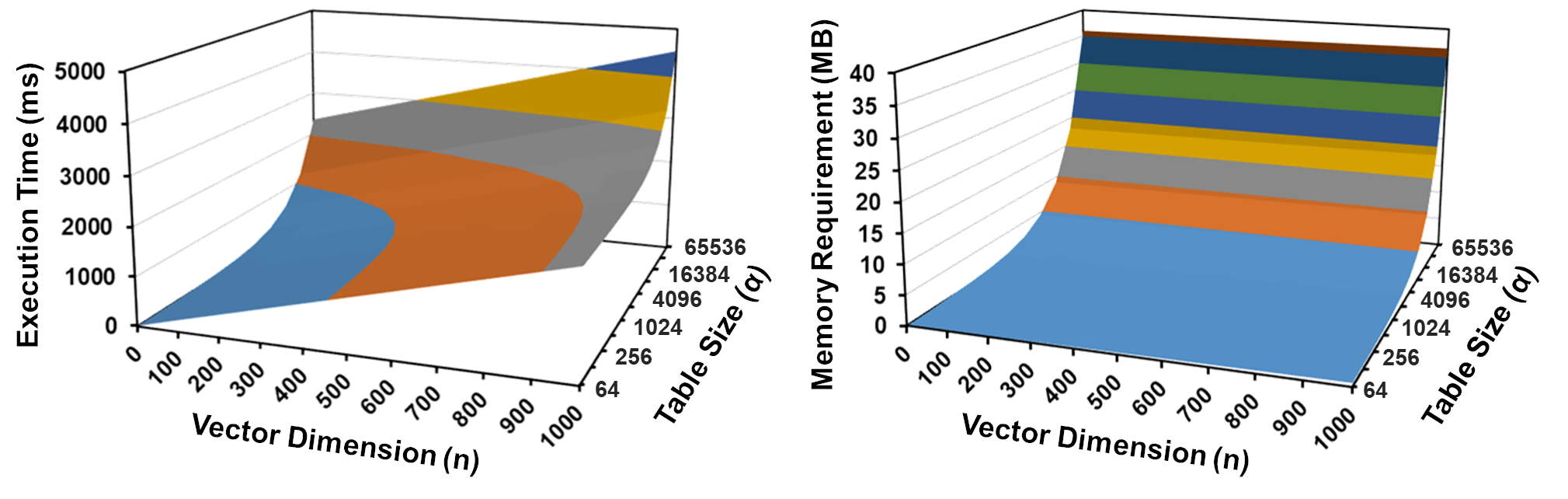}
\caption{Compute cost of FHIPE \textsf{Decrypt} for different vector dimensions ($n$) and table size ($\alpha$): (left) execution time and (right) memory requirement.}
\label{fig:fhipe_decrypt_cost}
\end{figure}

\begin{figure}[!t]
\centering
\includegraphics[width=3.4in]{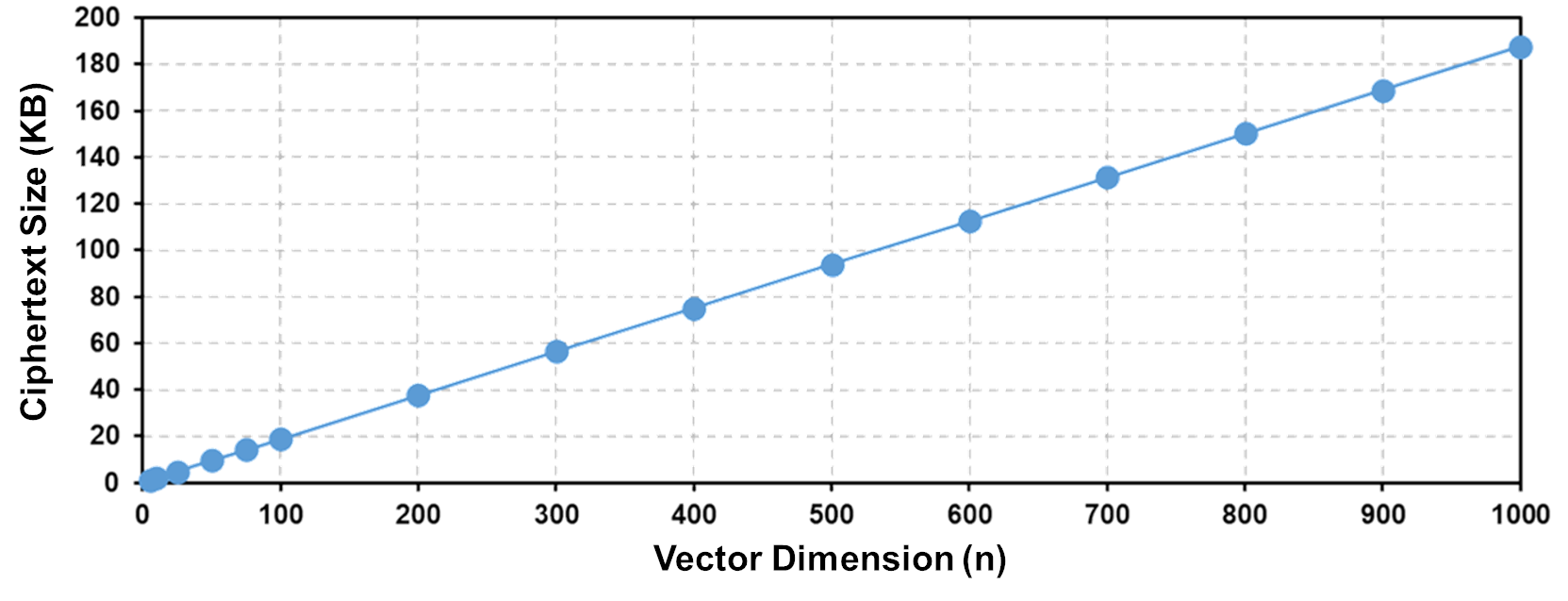}
\caption{FHIPE ciphertext size for different vector dimensions ($n$).}
\label{fig:ciphertext_size}
\end{figure}

The execution time of \textsf{Encrypt} is $O(n)$, dominated by the $n+1$ scalar multiplications in $\Gb$, while its memory requirement is $O(n^2)$ due to the storage of $n \times n$ matrix $\boldsymbol{B}^{*} \in \GLn(\Zq)$. This is confirmed by the measured results shown in Figure \ref{fig:fhipe_encrypt_cost}.
The execution time and memory requirement of \textsf{Decrypt} are both $O(n) + O(\alpha)$. The execution time is dominated by the $n$-fold multi-pairing for $n \gg \alpha$ and by the bounded discrete logarithm for $\alpha \gg n$. The memory requirement is dominated by the discrete logarithm lookup table for reasonably large $\alpha$. This is confirmed by the measured results shown in Figure \ref{fig:fhipe_decrypt_cost} (the $n$-axis is linear and the $\alpha$-axis is logarithmic to base 2).
The communication cost is $O(n)$ as the FHIPE ciphertext consists of $n+1$ points in $E'(\Fpa)$, that is, $4(n+1)$ $\Fp$ elements. This is shown in Figure \ref{fig:ciphertext_size}.

The above results are obtained by averaging over 10,000 runs with the optimized BLS12-381-based implementation.
Kim et al. \cite{kim_ipe_2018} had presented preliminary implementation results of their FHIPE scheme using the MNT6-224 pairing-friendly curve (with 224-bit prime field $\Fp$, embedding degree $k=6$ and $\approx$ 80-bit security level) on a 4.0~GHz octa-core Intel Core-i7 processor with 16~GB RAM.
Elliptic curve scalar multiplication and pairing operations become increasingly more computationally expensive with larger size of the underlying prime field. However, the optimized implementation proposed in this work achieves performance comparable with \cite{kim_ipe_2018} despite being implemented on an edge device with a slower processor and supporting a curve at much higher security level with a larger prime field, as shown in Table \ref{table:perf_comparison}.
The use of twist curve $E'$ representation of $\Gb$ also leads to significant compression of ciphertext size despite larger $p$.

\begin{table*}[!t]
\renewcommand{\arraystretch}{1.25}
\caption{Performance Comparison with Previous Work (for $n=100$ and $\alpha=1024$)}
\label{table:perf_comparison}
\centering
\begin{tabular}{|l|c|c|c|c|c|c|c|}
\hline
\rowcolor{Gray}
\textbf{Implementation} & \textbf{Curve} & \textbf{Prime Size} & \textbf{Security Level} & \textbf{Measurement Setup} & \textbf{Encrypt} & \textbf{Decrypt} & \textbf{Ciphertext} \\
\hline
Kim et al. \cite{kim_ipe_2018} & MNT6-224 & 224-bit & $\approx$ 80-bit & 4.0~GHz Intel Core-i7 (16~GB RAM) & 71.4~ms & 366.4~ms & 17.7~KB \\
\hline
\textbf{This work} & BLS12-381 & 381-bit & $\approx$ 126-bit & 1.8~GHz Raspberry Pi 4B (8~GB RAM) & 165~ms & 266~ms & 18.9~KB \\
\hline
\end{tabular}
\end{table*}
\section{Applications}
\label{sec:applications}

\begin{figure}[!t]
\centering
\includegraphics[width=3.25in]{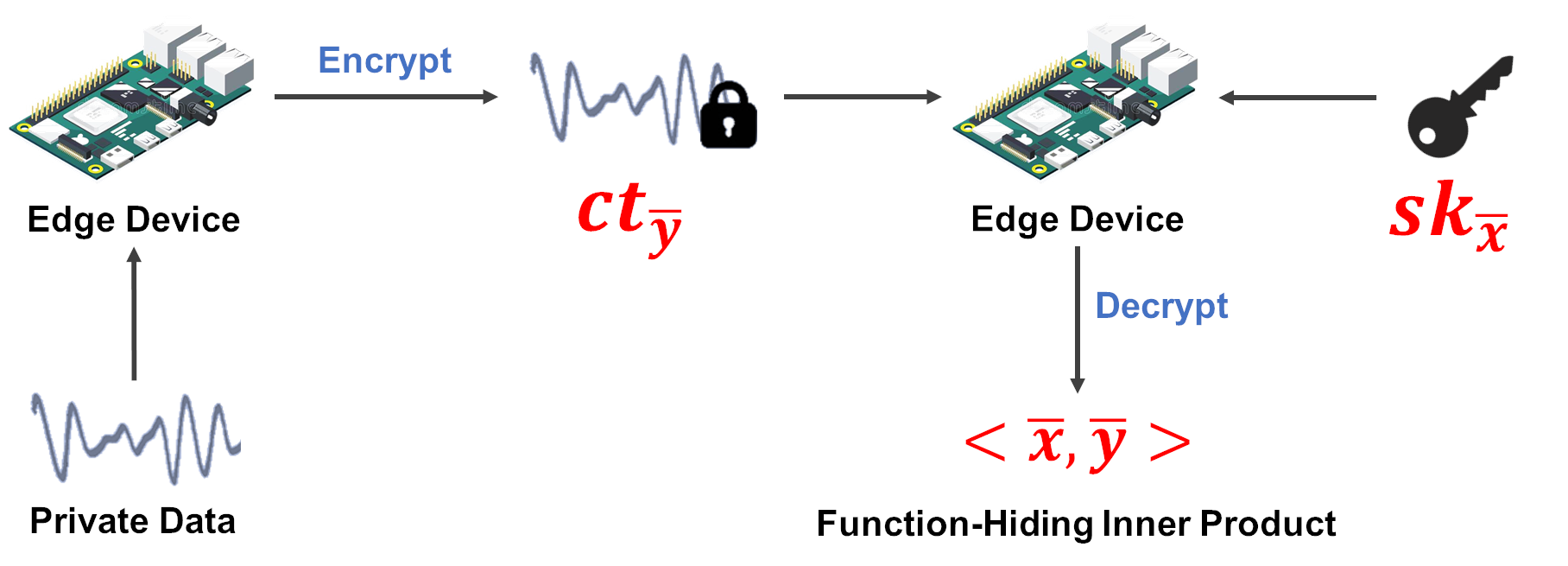}
\caption{Privacy-preserving inner product computation using FHIPE.}
\label{fig:fhipe_system}
\end{figure}

Some of the privacy-preserving computation applications which can benefit from this efficient implementation of FHIPE in IoT / edge devices are: (A) biomedical sensor data classification, and (B) wireless fingerprint-based indoor localization.
Figure \ref{fig:fhipe_system} shows a system diagram where the encryptor, an edge device, generates ciphertext $ct_{\boldsymbol{y}}$ (corresponding to private vector $\boldsymbol{y}$) and the decryptor, also an edge device, uses secret key $sk_{\boldsymbol{x}}$ (corresponding to private vector $\boldsymbol{x}$) to decrypt the ciphertext and obtain the inner product $\langle \boldsymbol{x}, \boldsymbol{y} \rangle$.
The FHIPE scheme naturally supports both negative and positive vector elements, where a negative input vector element $x < 0$ is mapped to modulo $q$ domain as $q + x$.
The set $S$ in baby-step giant-step algorithm can also be modified to support both positive and negative  exponents. If the bounded discrete logarithm result $z > \lceil \, q/2 \, \rceil$, it is mapped from modulo $q$ domain as $z - q$. Since $s \ll q$, assuming $s < \lceil \, q/2 \, \rceil$, positive and negative values always remain disjoint in the modulo $q$ domain.
The mapping of this framework to the above applications as well as their performance analysis are described next.

\subsection{Privacy-Preserving Biomedical Data Classification}

Linear classifiers \cite{gallant_perceptron_1990} for encrypted biomedical data are implemented using the FHIPE framework which preserves confidentiality of both input data and classifier weights.
The encryptor sends the ciphertext corresponding to the biomedical signal as $ct_{\boldsymbol{s}}$ = \textsf{Encrypt} ($msk$, $\boldsymbol{s}$).
The decryptor evaluates $z = \langle \boldsymbol{s}, \boldsymbol{w} \rangle$ = \textsf{Decrypt} ($pp$, $sk_{\boldsymbol{w}}$, $ct_{\boldsymbol{s}}$) using the decryption key $sk_{\boldsymbol{w}}$ = \textsf{KeyGen} ($msk$, $\boldsymbol{w}$) which embeds the classification model (weights).
The decryption results in classification to class $C_0$ (normal) or $C_1$ (abnormal) based on whether $z \le T$ or $z > T$ respectively, where the decryptor knows classification threshold $T$.
Neither input vector $\boldsymbol{s}$ nor weight vector $\boldsymbol{w}$ are revealed.

\begin{figure}[!t]
\centering
\includegraphics[width=2.9in]{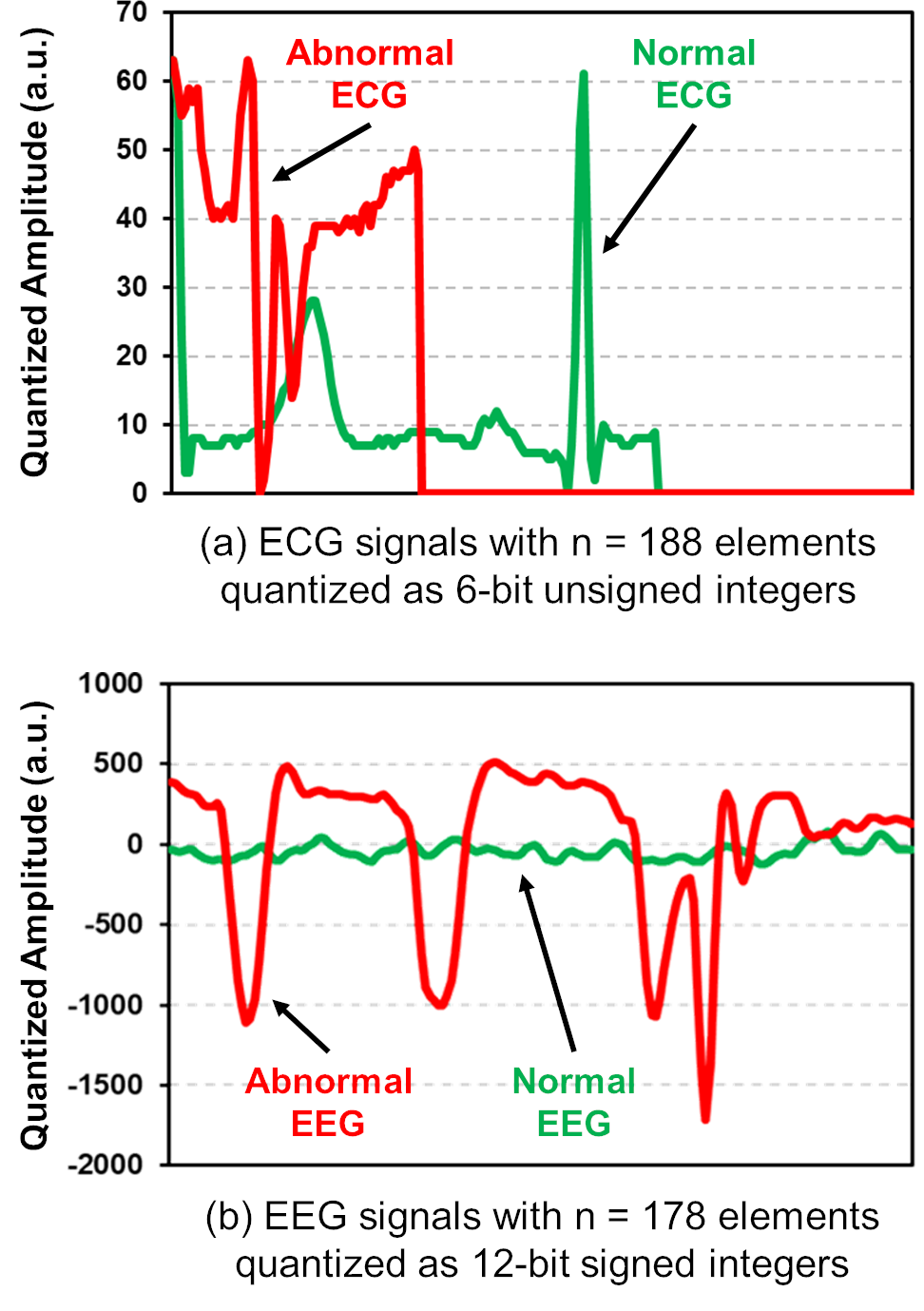}
\caption{Illustrative ECG and EEG data samples after quantization.}
\label{fig:biomedical_data}
\end{figure}

As practical examples, privacy-preserving linear classifiers for electrocardiogram (ECG) and electroencephalogram (EEG) signals are implemented with the open-source Physionet PTB Diagnostic ECG Database \cite{bousseljot_ecg_1995} and UCI Epileptic Seizure Recognition Dataset \cite{andrzejak_eeg_2001} respectively.
Illustrative samples of normal / abnormal ECG and EEG are shown in Figure \ref{fig:biomedical_data}.
Single-layer perceptron neural networks are trained over these datasets using gradient descent \cite{gallant_perceptron_1990} and the perceptron bias gives the linear classifier threshold $T$.
The ECG dataset contains 14,552 samples (4,046 in $C_0$ and 10,506 in $C_1$), where each sample is a 188-dimensional vector. The EEG dataset contains 11,500 samples (9,200 in $C_0$ and 2,300 in $C_1$), where each sample is a 178-dimensional vector. The ECG and EEG samples are quantized as 6-bit unsigned integers and 12-bit signed integers respectively.
In both cases, 500 samples from each class are used for cross-validation and the remaining are used for training.
Both classifiers achieve $\approx 80\%$ accuracy. In the FHIPE framework, the ECG classifier has $n = 188$ and $\alpha = 16384$ so that \textsf{Encrypt} and \textsf{Decrypt} take 307~ms and 1004~ms respectively in the Raspberry Pi 4B setup described earlier. Similarly, the EEG classifier has $n = 178$ and $\alpha = 16384$ so that \textsf{Encrypt} and \textsf{Decrypt} take 291~ms and 1567~ms respectively in the same setup.

\subsection{Privacy-Preserving Indoor Localization}

\begin{figure}[!t]
\centering
\includegraphics[width=2.7in]{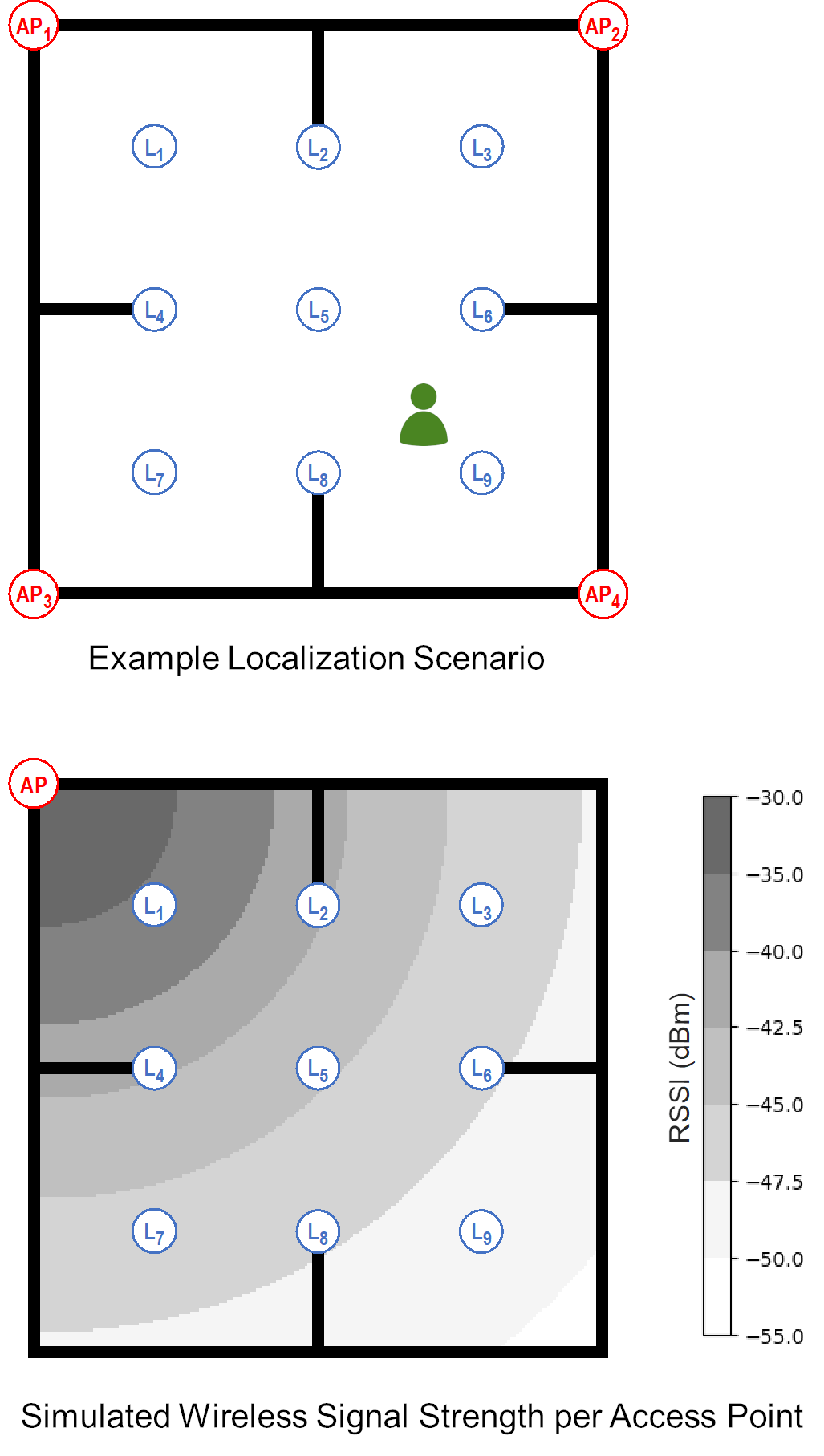}
\caption{Illustrative indoor localization scenario with $N=4$ access points $\{ AP_1, \cdots, AP_4 \}$ and $M=9$ database locations $\{ L_1, \cdots, L_9 \}$.}
\label{fig:localization_heatmap}
\end{figure}

Wireless fingerprint-based indoor localization \cite{li_localization_2014, jarvinen_localization_2018} is implemented using the FHIPE framework which preserves confidentiality of both client's input fingerprint and service provider's database.
Consider an indoor setting with $N$ wireless access points with unique public identifiers $AP_j \,\,\, \forall \,\,\, 1 \le j \le N$.
The set of Received Signal Strength Indicator (RSSI) values corresponding to all $N$ access points at a location $(x_i, y_i)$ together form the wireless fingerprint $\boldsymbol{v_i} = ( v_{i,1}, v_{i,2}, \cdots, v_{i,N} )$ at that location.
The setup phase involves the service provider creating a private database $D = \{ \, (i, \, (x_i, y_i), \, \boldsymbol{v_i} ) \, \}_{i=1}^{M}$ for $M$ locations of interest and a public the list of access point identifiers $T_{AP} = \{ \, AP_j \, \} _{j=1}^{N}$.
The operating phase involves the client measuring the wireless fingerprint $\boldsymbol{v} = ( v_{1}, v_{2}, \cdots, v_{N} )$ at its location, and the server computes squared Euclidean distances $d_i = \lVert \boldsymbol{v} - \boldsymbol{v_i} \rVert ^2 = \sum_{j=1}^{N} \, ( v_{j} - v_{i,j} )^2 = \sum_{j=1}^{N} \, v_{j}^2  + \sum_{j=1}^{N} \, (-2 \, v_{j} \, v_{i,j}) + \sum_{j=1}^{N} \, v_{i,j}^2$.
In the FHIPE setting, $M$ decryption keys are created corresponding to the database entries as $sk_{\boldsymbol{v'_{i}}}$ = \textsf{KeyGen} ($msk$, $\boldsymbol{v'_{i}} = ( \sum_{j=1}^{N} v_{i,j}^2, -2v_{i,1}, -2v_{i,2}, \cdots, -2v_{i,N}, 1)$) for $1 \le i \le M$.
The encryptor sends the client's encrypted fingerprint $ct_{\boldsymbol{v'}}$ = \textsf{Encrypt} ($msk$, $\boldsymbol{v'} = ( 1, v_{1}, v_{2}, \cdots, v_{N}, \sum_{j=1}^{N} v_{j}^2)$).
The decryptor evaluates $d_i = \lVert \boldsymbol{v} - \boldsymbol{v_i} \rVert ^2 = \langle \boldsymbol{v'}, \boldsymbol{v'_{i}} \rangle$ = \textsf{Decrypt} ($pp$, $sk_{\boldsymbol{v'_{i}}}$, $ct_{\boldsymbol{v'}}$) for $1 \le i \le M$, where $sk_{\boldsymbol{v'_{i}}}$ is the decryption key hiding the $i$-th database entry.
Location indices $i_1, i_2, \cdots \in [1, M]$ of the client's nearest neighbors are determined by these distances. The exact coordinates $(x_i, y_i)$ corresponding to the location indices $1 \le i \le M$ (including those of the nearest neighbours) remain secret throughout.
Neither client fingerprint $\boldsymbol{v}$ nor service provider database entries $\boldsymbol{v_i}$ are revealed.
Figure \ref{fig:localization_heatmap} shows an illustrative indoor localization scenario with $N = 4$ and $M = 9$, where the wireless fingerprint heat map is generated using the open-source tool from \cite{bicheru_wifiheatmap_2018}. The RSSI values (in dBm), which are all negative, are quantized as 6-bit unsigned integers. In the FHIPE framework, each of the $M$ distance computations has $n = N + 2 = 6$ and $\alpha = 64$ so that \textsf{Encrypt} and \textsf{Decrypt} take 12~ms and 29~ms respectively in the Raspberry Pi 4B setup described earlier.

\section{Conclusions and Future Work}
\label{sec:conclusion}

In this work, algorithmic optimizations and efficient software implementation of pairing-based function-hiding inner product encryption (FHIPE) \cite{kim_ipe_2018} have been presented using the recently proposed and widely adopted BLS12-381 pairing-friendly elliptic curve \cite{ietf_pairingcurves_2020}. Encryption performance is improved by $\approx 2.6\times$ using fast elliptic curve multi-scalar multiplication with GLS decomposition. Decryption performance is improved by $\approx 3.4\times$ using efficient sharing of Miller Loop and Final Exponentiation in  multi-pairing coupled with bounded discrete logarithm with power-tree-based lookup table generation. Point compression using twist curve representation provides $6\times$ reduction in ciphertext size.
Extensive performance analysis of the FHIPE implementation is provided using a Raspberry Pi 4B edge device. Privacy-preserving edge computing applications such as encrypted biomedical sensor data classification and secure wireless fingerprint-based indoor localization are also demonstrated using this setup.
The implementation framework presented in this work is based on the open-source MIRACL cryptographic software library \cite{miracl_crypto} and will be extended to cryptographic hardware acceleration \cite{banerjee_sscl_2021} in future work. Other privacy-preserving classification tasks such as logistic regression, support vector machine and multi-layer neural network will also be explored in the future.

\section*{Acknowledgment}

This work was supported by a Young Investigator Award from the Pratiksha Trust, Bangalore.


\balance
\bibliographystyle{IEEEtran}
\bibliography{references}

\begin{thebibliography}{10}
\providecommand{\url}[1]{#1}
\csname url@samestyle\endcsname
\providecommand{\newblock}{\relax}
\providecommand{\bibinfo}[2]{#2}
\providecommand{\BIBentrySTDinterwordspacing}{\spaceskip=0pt\relax}
\providecommand{\BIBentryALTinterwordstretchfactor}{4}
\providecommand{\BIBentryALTinterwordspacing}{\spaceskip=\fontdimen2\font plus
\BIBentryALTinterwordstretchfactor\fontdimen3\font minus \fontdimen4\font\relax}
\providecommand{\BIBforeignlanguage}[2]{{%
\expandafter\ifx\csname l@#1\endcsname\relax
\typeout{** WARNING: IEEEtran.bst: No hyphenation pattern has been}%
\typeout{** loaded for the language `#1'. Using the pattern for}%
\typeout{** the default language instead.}%
\else
\language=\csname l@#1\endcsname
\fi
#2}}
\providecommand{\BIBdecl}{\relax}
\BIBdecl

\bibitem{shan_securecompute_2018}
Z.~{Shan} \emph{et~al.}, ``{Practical Secure Computation Outsourcing: A Survey},'' \emph{ACM Computing Surveys (CSUR)}, vol.~51, no.~2, pp. 1--40, Feb. 2018.

\bibitem{menezes_handbook_2018}
A.~J. {Menezes}, P.~C. {Van Oorschot}, and S.~A. {Vanstone}, \emph{{Handbook of Applied Cryptography}}.\hskip 1em plus 0.5em minus 0.4em\relax CRC Press, 2018.

\bibitem{boneh_fe_2011}
D.~{Boneh}, A.~{Sahai}, and B.~{Waters}, ``{Functional Encryption: Definitions and Challenges},'' in \emph{Theory of Cryptography Conference (TCC)}, 2011, pp. 253--273.

\bibitem{kim_ipe_2018}
S.~{Kim} \emph{et~al.}, ``{Function-Hiding Inner Product Encryption Is Practical},'' in \emph{Security and Cryptography for Networks (SCN)}, 2018, pp. 544--562.

\bibitem{nadia_pairing_2017}
N.~{El Mrabet} and M.~{Joye}, \emph{{Guide to Pairing-Based Cryptography}}.\hskip 1em plus 0.5em minus 0.4em\relax CRC Press, 2017.

\bibitem{hankerson_ecc_2006}
D.~{Hankerson}, A.~{Menezes}, and S.~{Vanstone}, \emph{{Guide to Elliptic Curve Cryptography}}.\hskip 1em plus 0.5em minus 0.4em\relax Springer Science \& Business Media, 2006.

\bibitem{scott_ecciot_2020}
M.~{Scott}, ``{On the Deployment of Curve Based Cryptography for the Internet of Things},'' Cryptology ePrint Archive, Paper 2020/514, 2020.

\bibitem{ietf_pairingcurves_2020}
Y.~{Sakemi}, T.~{Kobayashi}, T.~{Saito}, and R.~S. {Wahby}, ``{Pairing-Friendly Curves},'' Crypto Forum Research Group (CFRG) Internet Draft, Internet Engineering Task Force (IETF), 2022, \url{https://tools.ietf.org/html/draft-irtf-cfrg-pairing-friendly-curves-11}.

\bibitem{raspberry_pi_4b}
{Raspberry Pi Foundation}, ``{Raspberry Pi 4B Single-Board Computer},'' \url{https://www.raspberrypi.org/products/raspberry-pi-4-model-b}.

\bibitem{miracl_crypto}
{MIRACL}, ``{MIRACL Core Cryptographic Library},'' https://github.com/miracl/core.

\bibitem{banerjee_phd_2021}
U.~{Banerjee}, ``{Efficient Algorithms, Protocols and Hardware Architectures for Next-Generation Cryptography in Embedded Systems},'' Ph.D. dissertation, Massachusetts Institute of Technology, 2021.

\bibitem{galbraith_endomorphism_2009}
S.~D. {Galbraith}, X.~{Lin}, and M.~{Scott}, ``{Endomorphisms for Faster Elliptic Curve Cryptography on a Large Class of Curves},'' in \emph{Advances in Cryptology (EUROCRYPT)}, 2009, pp. 518--535.

\bibitem{bos_exponentiating_2014}
J.~W. {Bos}, C.~{Costello}, and M.~{Naehrig}, ``{Exponentiating in Pairing Groups},'' in \emph{Selected Areas in Cryptography (SAC)}, 2014, pp. 438--455.

\bibitem{faz_efficient_2015}
A.~{Faz-Hern{\'a}ndez} \emph{et~al.}, ``{Efficient and Secure Algorithms for GLV-Based Scalar Multiplication and their Implementation on GLV-GLS Curves},'' \emph{Journal of Cryptographic Engineering}, vol.~5, pp. 31--52, 2015.

\bibitem{granger_product_2006}
R.~{Granger} and N.~P. {Smart}, ``{On Computing Products of Pairings},'' Cryptology ePrint Archive, Paper 2006/172, 2006.

\bibitem{scott_pairing_2019}
M.~{Scott}, ``{Pairing Implementation Revisited},'' Cryptology ePrint Archive, Paper 2019/077, 2019.

\bibitem{knuth_comp_1997}
D.~E. {Knuth}, \emph{{The Art of Computer Programming, Volume 2: Seminumerical Algorithms}}, 3rd~ed.\hskip 1em plus 0.5em minus 0.4em\relax Addison-Wesley, 1997.

\bibitem{ahsan_ants_2022}
F.~{Ahsan} and U.~{Banerjee}, ``{Embedded Software Implementation of Privacy Preserving Matrix Computation using Elliptic Curve Cryptography for IoT Applications},'' in \emph{IEEE International Conference on Advanced Networks and Telecommunications Systems (ANTS)}, 2022.

\bibitem{gallant_perceptron_1990}
S.~I. {Gallant}, ``{Perceptron-Based Learning Algorithms},'' \emph{IEEE Transactions on Neural Networks}, vol.~1, no.~2, pp. 179--191, 1990.

\bibitem{bousseljot_ecg_1995}
R.~{Bousseljot} \emph{et~al.}, ``{Nutzung der EKG-Signaldatenbank CARDIODAT der PTB {\"u}ber das Internet},'' \emph{Biomedical Engineering / Biomedizinische Technik}, vol.~40, no.~s1, pp. 317--318, 1995.

\bibitem{andrzejak_eeg_2001}
R.~G. {Andrzejak} \emph{et~al.}, ``{Indications of Nonlinear Deterministic and Finite-Dimensional Structures in Time Series of Brain Electrical Activity: Dependence on Recording Region and Brain State},'' \emph{Physical Review E}, vol.~64, no.~6, p. 061907, 2001.

\bibitem{li_localization_2014}
H.~{Li} \emph{et~al.}, ``{Achieving Privacy Preservation in WiFi Fingerprint-Based Localization},'' in \emph{IEEE Conference on Computer Communications (INFOCOM)}, 2014, pp. 2337--2345.

\bibitem{jarvinen_localization_2018}
Z.~{Yang} and K.~{Järvinen}, ``{The Death and Rebirth of Privacy-Preserving WiFi Fingerprint Localization with Paillier Encryption},'' in \emph{IEEE Conference on Computer Communications (INFOCOM)}, 2018, pp. 1223--1231.

\bibitem{bicheru_wifiheatmap_2018}
C.~{Bicheru}, ``{WiFi Heat Map Simulator},'' \url{https://github.com/cristian-bicheru/wifi-heatmap}.

\bibitem{banerjee_sscl_2021}
U.~{Banerjee} and A.~P. {Chandrakasan}, ``{A Low-Power BLS12-381 Pairing Cryptoprocessor for Internet-of-Things Security Applications},'' \emph{IEEE Solid-State Circuits Letters}, vol.~4, pp. 190--193, 2021.

\end{thebibliography}

\end{document}